# Effects of spontaneously generated coherence on resonance fluorescence from lateral triple quantum -dot molecules

Si-Cong Tian[a], Cun-Zhu Tong[a,*], Chun-Liang Wang[b], Yong-Qiang Ning[a], Li-Jun Wang[a]

[a]*State Key laboratory of Luminescence and Application, Changchun Institute of Optics Fine Mechanics and Physics Chinese Academy of Sciences, Changchun, 130033, PR China*

[b]*Centre for Advanced Optoelectronic Functional Materials Research and Key Laboratory for UV Light-Emitting Materials and Technology of Ministry of Education, Northeast Normal University, Changchun 130024, PR China*

*Corresponding author: tiansicong@ciomp.ac.cn (S.-C. Tian)

tongcz@ciomp.ac.cn (C.-Z. Tong)

**ABSTRACT**

We investigate the spectrum of the resonance fluorescence from the lateral triple quantum dots controlled by voltage and obtain some interesting features such as controllable triple narrow peaks. In our system we use tunneling instead of coupling lasers, and the positions, widths and heights of the resonance fluorescence peaks can be controlled by tuning the tunneling couplings. We explain the observed spectrum with the transition properties of the dressed states generated by the coupling of the two tunneling and the laser field. These features can also be viewed as the effects of Spontaneously Generated

Coherence between the close-lying levels in the dressed state picture of the tunneling couplings. And the scheme proposed here can permit the observation of Spontaneously Generated Coherence.

## 1. Introduction

Recently much interest has been attracted by the coherent manipulation of semiconductor quantum dots (QDs) [1-5]. This research is motivated by the proposal of using QDs as the basic building block of quantum information technology, such as qubits for quantum computers [6], single photon emitters [7], all-optical quantum gate [8] and entangled-photon sources [9]. In order to reduce decoherence effects, resonant optical excitation is often employed in experiments with semiconductor QDs. For an ideal two-level system, it is well known that strong resonant excitation with a continuous wave (cw) field results in a so-called resonance fluorescence (RF) spectrum containing three peaks, known as the Mollow triplet. This iconic spectral feature was first predicted by Mollow [10] and has been demonstrated experimentally for atoms [11], single molecules [12] and self-assembled semiconductor QDs [13,14]. The resonance fluorescence spectrum of multiple-level system driven by two or more lasers has much more structure and shows a variety of interesting effects, duo to the more complex internal dynamics of the system. And most recently, we have experimentally observed the resonance fluorescence spectrum for a driven cascade model system in the atomic beam [15].

On the other hand, we can modify the properties of resonance fluorescence spectrum by spontaneously generated coherence (SGC). SGC refers to the interference of two or more decay channels with nonorthogonal electric-dipole transition matrix elements. SGC can find applications in many fields such as quantum information and computation [16], lasing without inversion [17], all-optical switching [18], quantum photocell [19] and high-precision metrology [20]. In the presence of SGC, the spectrum of resonance fluorescence can be demonstrated to exhibit narrowing [21], quenching [22], squeezing [23] and superfluorescence [24]. But the existence of SGC requires two stringent conditions: the close-lying levels are near degenerate and the corresponding dipole transitions are not orthogonal. These conditions are very difficult to be fulfilled in real systems. As a result, there are hardly any direct experimental observations on the phenomena related to SGC.

A coupled triple quantum dots (TQDs) coupled by tunneling, is an artificial fully tunable molecule, where it is possible to create a multiple-level system. In such molecules, an external electric field allows us to control the confining potential and the number of electrons in each artificial atom, as well as their mutual interaction [25]. Interest in the coherent manipulation of TQDs is driven by the desire to develop capabilities to design increasingly complex quantum systems at both single-particle and many-particle level. Such a quantum system by design might serve as a laboratory for correlated electron systems as well as a prototype quantum processor based on charge and/or spin in QDs [26,27]. There are other applications of a TQDs in the area of quantum computation, such as quantum teleportation [28] and exchange-controlled qubit [29]. And also TQDs allows for application of tools known from quantum optics, such as the rectification [30] or the coherent electronic transfer using adiabatic passage [31].

In this paper, we investigate the resonance fluorescence spectrum from lateral linear TQDs. And much experimental progress has been achieved with such TQDs [32-35]. By the bias voltage in the TQDs, we obtain a few interesting features, such as triple narrow lines in the resonance fluorescence spectrum. Unlike the atomic system, in TQDs we can use tunneling instead of coupling lasers to create the necessary coherence, and the positions, linewidths and heights of the fluorescence peaks can be controlled. We explain the phenomena in the dressed state basis of the laser field and the tunneling couplings. On the other hand, in the dressed state basis of the two tunneling couplings, the scheme turns to be a four-level system with SGC. We can also attribute the observed spectrum of resonance fluorescence to the SGC between the three upper dressed levels. So the TQDs proposed here can give a feasible way to experimentally observe the effect of SGC, but without the need for near-degenerated levels and nonorthogonal dipole moments.

The remainder of this paper is organized as follows. In Section 2, we describe the TQD model and calculate resonance fluorescence spectrum of the TQD system. In Section 3, we discussed the numerical results of the resonance fluorescence spectrum. Section 4 contains a summary and outlook.

## 2. Model and dynamic equations

We show the schematic of the setup, band structure and level configuration of a TQD system in Fig. 1 (a). Without the excitation of the laser, no excitons are inside all QDs, which corresponds to state $|0\rangle$. When a laser field is applied, a direct exciton is created inside the QD 1, condition represented by the state $|1\rangle$. The band profiles alignment can be modified by the external electric field, in this case the electron can tunnel from QD 1 to the QD 2, and from QD 2 to QD 3. And we denote these indirect excitons as state $|2\rangle$ and state $|3\rangle$. In such system, we can controlled the tunnel barrier by placing a gate electrode between the neighboring dots.

In the interaction picture and the rotating wave and dipole approximations, the Hamiltonian of this system is (we use units such that $\hbar = 1$)

$$H = \sum_{j=0}^{3} E_j |j\rangle\langle j| + [(\Omega_c e^{-i\omega_c t}|0\rangle\langle 1| + T_1|2\rangle\langle 1| + T_2|3\rangle\langle 2|) + \text{H.c.}], \tag{1}$$

where $E_j = \hbar\omega_j$ is the energy of state $|j\rangle$, $\omega_c$ is the laser frequency, $\Omega_c = \boldsymbol{\mu}_{01} \cdot \mathbf{e} \cdot E$ is the Rabi frequency of the transition $|0\rangle \to |1\rangle$, with $\boldsymbol{\mu}_{01}$ being the associated dipole transition-matrix element, $\mathbf{e}$ the polarization vector and $E$ the electric-field amplitude of the laser pulse. And $T_1$ and $T_2$ are the tunneling coupling.

From the Liouville equation, we obtain the following equations for the density-matrix elements:

$$\dot{\rho}_{01} = i[\Omega_c(\rho_{11} - \rho_{00}) - T_1\rho_{02}] + (i\delta_1 - \Gamma_{10} - \gamma_1)\rho_{01}, \tag{2a}$$

$$\dot{\rho}_{02} = i(\Omega_c\rho_{12} - T_1\rho_{01} - T_2\rho_{03} -) + \frac{1}{2}[i(\delta_1 + \delta_2) - \Gamma_{20} - \gamma_2]\rho_{02}, \tag{2b}$$

$$\dot{\rho}_{03} = i(\Omega_c\rho_{13} - T_2\rho_{02}) + \frac{1}{2}[i(\delta_1 + \delta_3) - \Gamma_{30} - \gamma_3]\rho_{03}, \tag{2c}$$

$$\dot{\rho}_{11} = i[\Omega_c(\rho_{01} - \rho_{10}) + T_1(\rho_{21} - \rho_{12})] - \Gamma_{10}\rho_{11}, \tag{2d}$$

$$\dot{\rho}_{12} = i[T_1(\rho_{22} - \rho_{11}) + i\Omega_c\rho_{02} - T_2\rho_{13}] + \frac{1}{2}[i(\delta_1 - \delta_2) - \Gamma_{10} - \Gamma_{20} - \gamma_1 - \gamma_2]\rho_{12}, \tag{2e}$$

$$\dot{\rho}_{13} = i[\Omega_c\rho_{03} + T_1\rho_{23} - T_2\rho_{12}] + \frac{1}{2}[i(\delta_1 - \delta_3) - \Gamma_{10} - \Gamma_{30} - \gamma_1 - \gamma_3]\rho_{13}, \tag{2f}$$

$$\dot{\rho}_{22} = iT_1(\rho_{12} - \rho_{21}) + iT_2(\rho_{32} - \rho_{23}) - \Gamma_{20}\rho_{22}, \tag{2g}$$

$$\dot{\rho}_{23} = i[T_2(\rho_{33} - \rho_{22}) + T_1\rho_{13}] + \frac{1}{2}[i(\delta_2 - \delta_3) - \Gamma_{20} - \Gamma_{30} - \gamma_2 - \gamma_3]\rho_{23}, \tag{2h}$$

$$\dot{\rho}_{33} = iT_2(\rho_{23} - \rho_{32}) - \Gamma_{30}\rho_{33}, \tag{2i}$$

$$\dot{\rho}_{ij} = -\dot{\rho}_{ji}^{*}, \tag{2j}$$

$$\rho_{00} + \rho_{11} + \rho_{22} + \rho_{33} = 1. \tag{2k}$$

Here the detunings are defined as $\delta_1 = (\omega_{01} - \omega_c)$, $\delta_2 = \delta_1 + 2\omega_{21}$ and $\delta_3 = \delta_1 + 2\omega_{21} + 2\omega_{31}$, with $\omega_{mn}$ the transition frequency between $|m\rangle$ and $|n\rangle$ states. And $\Gamma_{10}$, $\Gamma_{20}$ and $\Gamma_{30}$ are the radiative decay rate of populations from $|1\rangle \to |0\rangle$, $|2\rangle \to |0\rangle$ and $|3\rangle \to |0\rangle$, and $\gamma_1$, $\gamma_2$ and $\gamma_3$ are the pure dephasing rate. And we denote $\Gamma_1 = \Gamma_{10} + \gamma_1$, $\Gamma_2 = \Gamma_{20} + \gamma_2$ and $\Gamma_3 = \Gamma_{30} + \gamma_3$ as the effective parameters.

Following the common method used in Ref. [15] and [21], we calculate the steady-state fluorescence spectrum. As is well known, the fluorescence spectrum is proportional to the Fourier transform of the steady-state correlation function $\lim_{t \to \infty} \langle E^{(-)}(r, \tau+t) \cdot E^{(+)}(r,t) \rangle$ [36], where $E^{(\pm)}(r,t)$ are the positive and negative frequency parts of the radiation field in the far zone, which consists of a free field operator, and a source-field operator that is proportional to the atomic polarization operator. Thus the incoherent fluorescence spectrum $S(\omega)$ can be expressed in terms of the atomic correlation function

$$S(\omega) = \mathrm{Re} \int_0^\infty \lim_{t \to \infty} \langle \Delta D^+(\tau+t) \cdot \Delta D^-(t) \rangle e^{-i\omega\tau} d\tau, \tag{3}$$

where Re denotes the real part, and $\Delta D^\pm(t) = \Delta D^\pm(t) - \langle \Delta D^\pm(\infty) \rangle$ is the deviation of the dipole polarization operator $D^\pm(t)$ from its mean steady-state value, and

$$D^+(t) = \mu_{01} a_1^\dagger a_2, \tag{4a}$$

$$D^-(t) = [D^+(t)]^\dagger, \tag{4b}$$

Then we rewrite the Eq. (2) in the form

$$\frac{d}{dt}\Psi = \mathbf{L}\Psi + \mathbf{I}, \tag{5}$$

where $\Psi = (\rho_{01}, \rho_{02}, \rho_{03}, \rho_{10}, \rho_{11}, \rho_{12}, \rho_{13}, \rho_{20}, \rho_{21}, \rho_{22}, \rho_{23}, \rho_{30}, \rho_{31}, \rho_{32}, \rho_{33})^T$, and $\mathbf{L}$ is a $15 \times 15$ matrix. The elements of $\mathbf{L}$ and $\mathbf{I}$ can be found explicitly from Eq. (2).

By means of the quantum regression theorem [37,38], the steady spectrum of the fluorescence spectrum from the state $|1\rangle$ to the ground state $|0\rangle$ can be obtained

$$S(\delta_k) = \text{Re}\{M_{11}\bar{\rho}_{11} + M_{12}\bar{\rho}_{12} + M_{13}\bar{\rho}_{13} + \sum_l N_{1l} I_l \bar{\rho}_{10}\}, \tag{6}$$

where

$$M_{ij} = [(z-L)^{-1}|_{z=i\delta_k}]_{i,j}, \quad N_{ij} = [L^{-1}(z-L)^{-1}|_{z=i\delta_k}]_{i,j}. \tag{7}$$

$\bar{\rho}_{ij}$ is the steady-state population ($i = j$) and atomic coherence ($i \neq j$), which can be obtained by setting $\dot{\rho}_{ij} = 0$ and solving numerically Eq. (2). $\delta_k$ is the detuing between the fluorescence and the transition $|1\rangle \rightarrow |0\rangle$. In the next part, we will calculate the corresponding fluorescence spectrum by tuning the coupling of the tunneling. And this parameter depends on the barrier characteristics and the external electric field.

## 3. Results and discussions

For our investigation, we assume that the radiative decay rate of populations $\Gamma_{10} \simeq 6.6\mu\text{eV}$ [39], $\Gamma_{20} = \Gamma_{30} \simeq 10^{-4}\Gamma_{10}$ [40], and the effective dephasing parameters, $\Gamma_1 \simeq 20\mu\text{eV}$ [41], $\Gamma_2 = \Gamma_3 \simeq 10^{-3}\Gamma_1$ [42]. The tunneling couplings can be controlled by the barrier characteristics and the external electric field. We mainly focus on the case of $\delta_1 = \omega_{21} = \omega_{32} = 0$.

First, we fix the value of $\Omega_c$ and $T_1$ to $20\mu\text{eV}$, and obtain the results in Fig. 2 by changing the value of $T_2$. When $T_2 = 0$, which means the electron can tunnel from QD 1 to the QD 2, but can not tunnel from QD 2 to QD 3. Thus the system is like the double QD system and five bored peaks appear in the spectrum [Fig. 2(a)]. The height of the middle peak is lower than the inner bands. When the tunneling coupling $T_2$ is increased to $T_2 = 4\mu\text{eV}$, which means three QDs are coupled together by the two tunneling couplings, we can see that three narrow peaks appear in the middle of the spectrum. And the height of these middle peaks is larger than the sidebands [Fig. 2(b)]. Then when $T_2 = 8\mu\text{eV}$, the intensity of the three narrow peaks in the middle is increased, as well as the splittings between these peaks [Fig. 2(c)]. When we increase the tunneling coupling to $T_2 = 16\mu eV$, the width of the three peaks in the middle are in the same order as other peaks, and the height of the peaks continue to increase [Fig. 2(d)].

It is well known that positions, heights, and widths of the fluorescence peaks are determined by the energies, steady-state populations, and electronic dipole moments of dressed states. In order to interpret the above numerical results, we investigate the properties of the dressed levels. Under the resonant coupling of the laser field and the two tunneling couplings, the energy eigenvalues of these dressed states [43] can be written as follows:

$$\lambda_4 = -\lambda_1 = \kappa_+ , \tag{8a}$$

$$\lambda_3 = -\lambda_2 = \kappa_- , \tag{8b}$$

where

$$\kappa_\pm = \sqrt{\frac{(\Omega_c^2 + T_1^2 + T_2^2) \pm \sqrt{(\Omega_c^2 + T_1^2 + T_2^2)^2 - 4\Omega_c^2 T_2^2}}{2}} . \tag{9}$$

Then the dressed levels can be express as

$$|\Psi_1\rangle = C_{10}|0\rangle + C_{11}|1\rangle + C_{12}|2\rangle + C_{13}|3\rangle , \tag{10a}$$

$$|\Psi_2\rangle = -C_{20}|0\rangle + C_{21}|1\rangle - C_{22}|2\rangle + C_{23}|3\rangle , \tag{10b}$$

$$|\Psi_3\rangle = -C_{30}|0\rangle + C_{31}|1\rangle - C_{32}|2\rangle + C_{33}|3\rangle , \tag{10c}$$

$$|\Psi_4\rangle = C_{40}|0\rangle + C_{41}|1\rangle + C_{42}|2\rangle + C_{43}|3\rangle , \tag{10d}$$

where

$$C_{i0} = \frac{1}{D_i} \frac{\Omega_c (\lambda_i^2 - T_2^2)}{T_1 T_2 \lambda_i} , \tag{11a}$$

$$C_{i1} = \frac{1}{D_i} \frac{\lambda_i^2 - T_2^2}{T_1 T_2} , \tag{11b}$$

$$C_{i2} = \frac{1}{D_i} \frac{\lambda_i}{T_2}, \tag{11c}$$

$$C_{i3} = \frac{1}{D_i}, \tag{11d}$$

$$D_i = \sqrt{1 + \left(\frac{\lambda_i}{T_2}\right)^2 + \left(\frac{\lambda_i^2 - T_2^2}{T_1 T_2}\right)^2 + \left(\frac{\Omega_c \left(\lambda_i^2 - T_2^2\right)}{T_1 T_2 \lambda_i}\right)^2}. \tag{11e}$$

And Eq. (10) can be written in the form

$$|\Psi_i\rangle = \sum C_{ik} |k\rangle, \text{ (i=a,b,c,d; k=0,1,2,3)} \tag{12}$$

So we can see that both the state $|1\rangle$ and state $|0\rangle$ [see Fig. 1(a)] are split into four dressed levels [see Fig. 1(b)], which are $|\Psi_1\rangle$, $|\Psi_2\rangle$, $|\Psi_3\rangle$ and $|\Psi_4\rangle$ for the bare-state level $|1\rangle$, while $|\Psi_1'\rangle$, $|\Psi_2'\rangle$, $|\Psi_3'\rangle$ and $|\Psi_4'\rangle$ for the bare-state level $|0\rangle$. Therefore the fluorescence from the state $|1\rangle$ to the ground state $|0\rangle$ has sixteen dipole transitions in the dressed-state representation. And every dipole transition is numbered in Fig. 1(b). Note that though $|\Psi_i\rangle$ and $|\Psi_i'\rangle$ are different in constant energy by the energy difference between level $|1\rangle$ and $|0\rangle$, the dressed levels $|\Psi_i\rangle$ has the same expressions and the eigenvalues as the dressed levels $|\Psi_i'\rangle$ with $i = 1, 2, 3, 4$.

In order to give an explicit explanation of the fluorescence spectrum, we fix the Rabi frequencies of the laser field $\Omega_c$ and the tunneling coupling $T_1$ at $\Omega_c = T_1 = 20\mu\text{eV}$, and show the dressed-state eigenvalues and corresponding populations as functions of the other tunneling coupling $T_2$ [see Fig. 3(a) and 3(b)]. It is obvious that the eigenvalues $\lambda_1$ and $\lambda_4$ depend weakly on $T_2$, while $\lambda_2$ and $\lambda_3$ are significantly affected by $T_2$. The populations of $|\Psi_1\rangle$, $|\Psi_2\rangle$, $|\Psi_3\rangle$ and $|\Psi_4\rangle$ are also influenced by $T_2$, and all the four dressed levels are well populated.

As we know that the decay rate of the transition between the dressed levels $|\Psi_i\rangle$ to $|\Psi_j'\rangle$ is proportional to the squared dipole moments $R_{ij} = |\langle j|\mathbf{P}|i\rangle|^2$, where $\mathbf{P} = \boldsymbol{\mu}_{01}|0\rangle\langle 1|$ is the transition dipole moment operator between $|0\rangle$ and $|1\rangle$ in the bare state basis. $R_{ij}$ can be calculated with the expression

$$R_{ij} = |\langle j|\mathbf{P}|i\rangle|^2 = |C_{j0}|^2 \mu_{01}^2 |C_{i1}|^2, \quad (i, j = 1, 2, 3, 4) \tag{13}$$

We can interpret the numerical results in Fig. 2 with the energies, populations and decay rates of the dressed levels. The triple peaks in the middle originate from the transitions 6, 7, 10 and 11 [see Fig. 2(b) and 2(c)]. When the tunneling coupling $T_2$ takes a small value, the decay from the levels $|\Psi_2\rangle$ and $|\Psi_3\rangle$ are much slower than that from $|\Psi_1\rangle$ and $|\Psi_4\rangle$. Therefore the triple peaks in the middle are much narrower than the others. The energy space, populations and decay rates of the dressed levels $|\Psi_2\rangle$ and $|\Psi_3\rangle$ grow with an increase of $T_2$, which results in the increase of splittings between the peaks, the heights and widths of the triple peaks in the middle.

The above results can also be viewed as the effect of SGC between the dressed levels. In the dressed-state representation of the tunneling coupling $T_1$ and $T_2$, the system turns to be a four-level scheme with three close-lying excited levels $|\Psi_0\rangle$, $|\Psi_+\rangle$ and $|\Psi_-\rangle$ [see Fig. 1(c)]. Under the resonant case, the energy eigenvalues of these dressed levels are $\lambda_0 = 0$ and $\lambda_\pm = \pm\sqrt{T_1^2 + T_2^2}$, and the corresponding eigenstates are

$$|\Psi_0\rangle = -\frac{T_2}{T}|1\rangle + \frac{T_1}{T}|3\rangle, \tag{14a}$$

$$|\Psi_+\rangle = \frac{1}{\sqrt{2}}\left(\frac{T_1}{T}|1\rangle - |2\rangle + \frac{T_2}{T}|3\rangle\right), \tag{14b}$$

$$|\Psi_-\rangle = \frac{1}{\sqrt{2}}\left(\frac{T_1}{T}|1\rangle + |2\rangle + \frac{T_2}{T}|3\rangle\right), \tag{14c}$$

where

$$T = \sqrt{T_1^2 + T_2^2}.  \qquad (15)$$

It is clear that the properties of the dressed levels, take energy positions as an example, are determined by the tunneling coupling $T_1$ and $T_2$. These three dressed levels decay to the ground level $|0\rangle$ with rates $\gamma_0 = \Gamma_{10} T_2^2 / T^2$, $\gamma_+ = \gamma_- = \Gamma_{10} T_1^2 / 2T^2$, respectively, and SGC arises from these three dressed levels. When $T_2$ is relatively small compared with $T_1$ and $\Omega 2$, the SGC between the three spontaneous emission pathways $|\Psi_0\rangle \to |0\rangle$, $|\Psi_+\rangle \to |0\rangle$ and $|\Psi_-\rangle \to |0\rangle$ are so strong that sharp lines arises in the fluorescence spectrum [see Fig. 2(b) and 2(c)]. When $T_2$ is large, the SGC between the spontaneous emission pathways are weaker, thus no linewidth narrowing occurs [see Fig. 2(d)].

From Eq. (14), the properties of the dressed levels $|\Psi_0\rangle$, $|\Psi_+\rangle$ and $|\Psi_-\rangle$ also depend on the tunneling coupling $T_1$. As a result, we can also anticipate the control of resonance fluorescence by changing the value of $T_1$. The numerical results are shown in Fig. 4, where the tunneling coupling $T_2$ is fixed at $T_2 = 10\mu eV$. We can also analyze the fluorescence spectrum in the dressed state picture of the three driving fields [see Fig. 1(b)]. First, when $T_1 = 0$, the electron can not tunnel from QD 1 to the QD 2 or from QD 2 to QD 3, which is like the case of single QD system. We can obtain Mollow-type resonance fluorescence spectrum shown in Fig. 4(a). When $T_1 \neq 0$, we can see that three narrow peaks appear in the resonance fluorescence spectrum. With increasing value of $T_1$, the intensity of the three narrow peaks in the middle and the splittings between these peaks are increasing [Fig. 4(b)-4(d)]. Similar to the case in Fig. 2, the triple peaks in the middle are induced by the four transitions 6, 7, 10 and 11. The decay rates from the dressed levels $|\Psi_2\rangle$ and $|\Psi_3\rangle$ decrease with a larger $T_1$. Therefore, we can obtain three narrow peaks in the spectrum.

Keeping the properties of the dressed states $|\Psi_0\rangle$ and $|\Psi_\pm\rangle$ unchanged, we calculate the resonance fluorescence spectrum by tuning the Rabi frequency of the laser field. We assume $T_1 = T_2 = 10\mu eV$, and calculate the resonance fluorescence spectrum with different $\Omega_c$. The results are shown in Fig. 5. When $\Omega_c$ is relatively small compared to that of the two tunneling coupling, i.e. $\Omega_c = 1\mu eV$, we can see that the spectrum consists of three peaks whose widths are of the same order [see Fig. 5(a)]. If the three couplings are of the same value $\Omega_c = T_1 = T_2 = 10\mu eV$, we obtain more peaks and the middle peaks are higher than the sidebands [see Fig. 5(b)]. When we increase the Rabi frequency to $\Omega_c = 20\mu eV$, nine peaks are obtained and three of them acquire linewidth narrowing. And the central peak becomes much larger than the other peaks [see Fig. 5(c)]. When $\Omega_c$ is large enough, the nine emission peaks are well resolved [see Fig. 5(d)], and the narrowing of the three peaks are very prominent.

The origin of the spectrum is clear in the dressed state basis of all couplings [see Fig. 1(b)]. In the case of $\Omega_c = 1\mu eV$, the levels $|\Psi_2\rangle$ and $|\Psi_3\rangle$ (also $|\Psi_2'\rangle$ and $|\Psi_3'\rangle$) are near degenerate, and we can only see three broad peaks [see Fig. 5(a)]. The decay of the transitions 1 and 16 are too weak to be observed, while the other 14 transitions shown in Fig. 1(b) contribute to the emission peaks. When $\Omega_c$ is increased to $10\mu eV$ [see Fig. 5(b)], we can see the contribution from all the 16 transitions. As the decay rates from the dressed levels $|\Psi_2\rangle$ and $|\Psi_3\rangle$ are of the order of spontaneously decay rate, we get only broad peaks. When we further increase the value of $\Omega_c$, the decay rates from the dressed levels $|\Psi_2\rangle$ and $|\Psi_3\rangle$ decreases. Therefore, we observe narrow lines in the resonance fluorescence spectrum [see Fig. 5(c) and 5(d)].

In the above discussions, we consider the case of $\omega_{21} = \omega_{32} = 0$. If $\omega_{21} \neq 0$, $\omega_{32} \neq 0$, it is difficult to write the expressions of the dressed levels in simple forms. While we can qualitatively infer that the energies, populations, and decaying rates of the dressed levels are modified, the fluorescence spectrum changes accordingly. We show two representative examples in Fig. 6. The nonzero value of $\omega_{21}$ and $\omega_{32}$ lead to the asymmetry of the spectrum and break the degeneracy of the transitions between the dressed levels. As a matter of fact, we can get as much as thirteen peaks in the fluorescence spectrum (no shown here) with proper parameters.

## 4. Conclusions and outlook

We have investigated the steady-state spectrum of resonant fluorescence from a TQD system. The spectrum can be controlled by the two tunneling couplings and some interesting features such as controllable triple narrow peaks has been obtained. We have analyzed the resonance fluorescence spectrum in the dressed state basis of the two tunneling couplings and the laser field. The phenomena can also be attributed to SGC between the close-lying levels in the dressed state picture of the two tunneling couplings.

As is well known, it is very difficult to realize SGC in real atomic system owing to the rigorous requirements. But in the TQD system, we can simulate SGC by the tunneling coupling. Thus by observation of linewidth narrowing of the fluorescence spectrum from the TQDs, the existence of SGC can be proved without the need for closely lying levels and parallel dipole moments. Such a system may open the way to study the SGC effect in solid medium, and can find applications in many fields such as lasing without inversion, quantum information and computation. And also the narrowing of the peaks can be used in high-precision metrology.

**Acknowledgment**

This work has been supported the financial support from the National Basic Research Program of China (Grant No. 2013CB933300), the National Natural Science Foundation of China (Grant No. 61076064, 61176046 and 11304308), and the Hundred Talents Program of Chinese Academy of Sciences.

**References**

[1]  X. Xu, B. Sun, P. R. Berman, D. G. Steel, A. S. Bracker, D. Gammon, and L. J. Sham, Science 317 (2007) 929.

[2]  A. Muller, E. B. Flagg, P. Bianucci, X. Y. Wang, D. G. Deppe, W. Ma, J. Zhang, G. J. Salamo, M. Xiao, and C. K. Shih, Phys. Rev. Lett. 99 (2007) 187402.

[3]  H. S. Borges, L. Sanz, J. M. Villas-Bôas, and A. M. Alcalde, Phys. Rev. B 81 (2010) 075322.

[4]  J. E. Rolon and S. E. Ulloa, Phys. Rev. B 82 (2010) 115307.

[5]  A. J. Ramsay and Semicond. Sci. Technol. 25 (2010) 103001.

[6]  D. Loss and D. P. DiVincenzo, Phys. Rev. A 57 (1998) 120.

[7]  P. Michler, A. Imamolu, M. D. Mason, P. J. Carson, G. F. Strouse, and S. K. Buratto, Nature 406 (2000) 968.

[8]  X. Q. Li, Y. W. Wu, D. Steel, D. Gammon, T. H. Stievater, D. S. Katzer, D. Park, C. Piermarocchi, and L. J. Sham, Science, 301 (2003) 809.

[9]  N. Akopian, N. H. Lindner, E. Poem, Y. Berlatzky, J. Avron, and D. Gershoni, Phys. Rev. Lett. 96 (2006) 130501.

[10] B. R. Mollow, "Power spectrum of light scattered by two-level systems," Phys. Rev.

188(5) (1969) 1969.

[11] F. Schuda, C. R. Stroud, Jr., and M. Hercher, J. Phys. B 7 (1974) L198.

[12] G. Wrigge, I. Gerhardt, J. Hwang, G. Zumofen, and V. Sandoghdar, Nature Phys. 4 (2008) 60.

[13] E. B. Flagg, A. Muller, J. W. Robertson, S. Founta, D. G. Deppe, M. Xiao, W. Ma, G. J. Salamo, and C. K. Shih, Nature Phys. 5 (2009) 203.

[14] A. N. Vamivakas, Y. Zhao, C. Y. Lu, and M. Atatüre, Nature Phys. 5 (2009) 198.

[15] S. C. Tian, C. L. Wang, C. Z. Tong, L. J. Wang, H. H. Wang, X. B. Yang, Z. H. Kang, and J. Y. Gao, Opt. Exp. 20 (2012) 23559.

[16] C. H. Bennet and D. P. Divincenzo, Nature 404 (2000) 247.

[17] J. H. Wu and J. Y. Gao, Phys. Rev. A 65 (2002) 063807.

[18] M. Bajcsy, S. Hofferberth, V. Balic, T. Peyronel, M. Hafezi, A. S. Zibrov, V. Vuletic, and M. D. Lukin, Phys. Rev. Lett. 102 (2009) 203902.

[19] M. O. Scully, Phys. Rev. Lett. 104 (2010) 207701.

[20] O. Postavaru, Z. Harman, and C. H. Keitel, "Phys. Rev. Lett. 106 (2011) 033001.

[21] P. Zhou and S. Swain, Phys. Rev. Lett. 77 (1996) 3995.

[22] F. L. Li and S. Y. Zhu, Phys. Rev. A 59 (1999) 2330.

[23] F. L. Li, S. Y. Gao, and S. Y. Zhu, Phys. Rev. A 67 (2003) 063818.

[24] M. A. Macovei, J. Phys. B 40 (2007) 387.

[25] C. Hsieh, Y. Shim, M. Korkusinski, and P. Hawrylak, "Rep. Prog. Phys. 75 (2012) 114501.

[26] J. A. Brum and P. Hawrylak, Superlatt. Microstruct. 22 (1997) 431.

[27] Daniel Loss and David P. DiVincenzo, Phys. Rev. A 57 (1998) 120.

[28] M. A.Nielsen, E. Knill and R. Laflamme, Nature 396 (1998) 52.

[29] D. P. DiVincenzo, D. Bacon, J. Kempe, G. Burkard, and K. B. Whaley, Nature 408

(2000) 339.

[30] M. Stopa, Phys. Rev. Lett. 88 (2002) 146802.

[31] Andrew D. Greentree, Jared H. Cole, A. R. Hamilton, and Lloyd C. L. Hollenberg, Phys. Rev. B 70 (2004) 235317.

[32] E. A. Laird, J. M. Taylor, D. P. DiVincenzo, C. M. Marcus, M. P. Hanson, and A. C. Gossard, Phys. Rev. B 82 (2010) 075403.

[33] S. Amaha, T. Hatano, H. Tamura, S. Teraoka, T. Kubo, Y. Tokura, D. G. Austing and S. Tarucha, Phys. Rev. B 85 (2012) 081301.

[34] L. Gaudreau, A. Kam, G. Granger, S. A. Studenikin, P. Zawadzki and A. S. Sachrajda Appl. Phys. Lett. 95 (2009) 193101.

[35] L. Gaudreau, G. Granger, A. Kam, G. Aers, S. Studenikin, P. Zawadzki, M. Pioro-Ladriere, Z. Wasilewski, and A. Sachrajda, Nature Phys. 8 (2012) 54.

[36] M. O. Scully and M. S. Zubairy, *Quantum Optics* (Cambridge University London Press, 1997), Chap. 10.

[37] M. Lax, Phys. Rev. 172 (1968) 350.

[38] S. Swain, J. Phys. A 14 (1981) 2577.

[39] Pochung Chen, C. Piermarocchi, and L. J. Sham, Phys. Rev. Lett. 87 (2001) 067401.

[40] V. Negoita, D.W. Snoke, K. Eberl, Phys. Rev. B 60 (1999) 2661.

[41] P. Borri, W. Langbein, U. Woggon, M. Schwab, and M. Bayer, Phys. Rev. Lett. 91 (2003) 267401.

[42] L. V. Butov, A. Zrenner, G. Abstreiter, G. Böhm, and G. Weimann, Phys. Rev. Lett. 73 (1994) 304.

[43] C. Cohen-Tannoudji, J. Dupont-Roc, and G. Grynberg, *Atom-Photon Interactions* (Wiley-VCH, Weinheim, 2004).

**Figure Captions**

Fig. 1. (a) Left: Schematic of the setup. An optical pulse transmits the left QD. V is a bias voltage. Right: Schematic band structure and level configuration of a linear TQD system. (b) Dressed-state under the coupling of the laser field and two tunneling. And dipole transitions are numbered from 1 to 16. (c) Dressed-state under the coupling of the two tunneling. And the interference occurs from the three decay channels.

Fig. 2. Calculated spectrum of resonance fluorescence with fixed value of $T_1 = 20\mu eV$ and different value of $T_2$: (a) $T_2 = 0$, (b) $T_2 = 4\mu eV$, (c) $T_2 = 8\mu eV$, (d) $T_2 = 16\mu eV$. The other parameters are $\Omega_c = 20\mu eV$, $\delta_1 = 0$, $\omega_{21} = \omega_{32} = 0$, $\Gamma_{10} = 6.6\mu eV$, $\Gamma_{20} = \Gamma_{30} = 10^{-4}\Gamma_{10}$, $\Gamma_1 = 20\mu eV$, $\Gamma_2 = \Gamma_3 = 10^{-3}\Gamma_1$. The number in Fig. 2 denotes which transition in Fig. 1(b) the peaks originate from.

Fig. 3. Properties of dressed levels as functions of the tunneling coupling $T_2$: (a) the eigen energies $\lambda_i$, (b) steady-state populations of the dressed state $|\Psi_i\rangle$, (c) decay rates of the dressed state $|\Psi_2\rangle$ and $|\Psi_3\rangle$. Other parameters are the same as those in Fig. 2.

Fig. 4. Calculated spectrum of resonance fluorescence with fixed value of $T_2 = 10\mu eV$ and different value of $T_1$: (a) $T_1 = 0$, (b) $T_1 = 10\mu eV$, (c) $T_1 = 20\mu eV$, (d) $T_1 = 30\mu eV$. Other parameters are the same as those in Fig. 2.

Fig. 5. Calculated spectrum of resonance fluorescence with fixed value of $T_1 = 20\mu eV$, $T_2 = 20\mu eV$, and different value of $\Omega_c$: (a) $\Omega_c = 1\mu eV$, (b) $\Omega_c = 10\mu eV$, (c) $\Omega_c = 20\mu eV$, (d) $\Omega_c = 30\mu eV$. Other parameters are the same as those in Fig. 2.

Fig. 6. Calculated spectrum of resonance fluorescence with fixed value of $T_1 = 20\mu eV$, $T_2 = 20\mu eV$, $\Omega_c = 10\mu eV$, and different value of $\omega_{21}$ and $\omega_{32}$: (a) $\omega_{21} = 0$, $\omega_{32} = 5\mu eV$, (b) $\omega_{21} = -5\mu eV$, $\omega_{32} = 10\mu eV$. Other parameters are the same as those in Fig. 2.

**Figures**

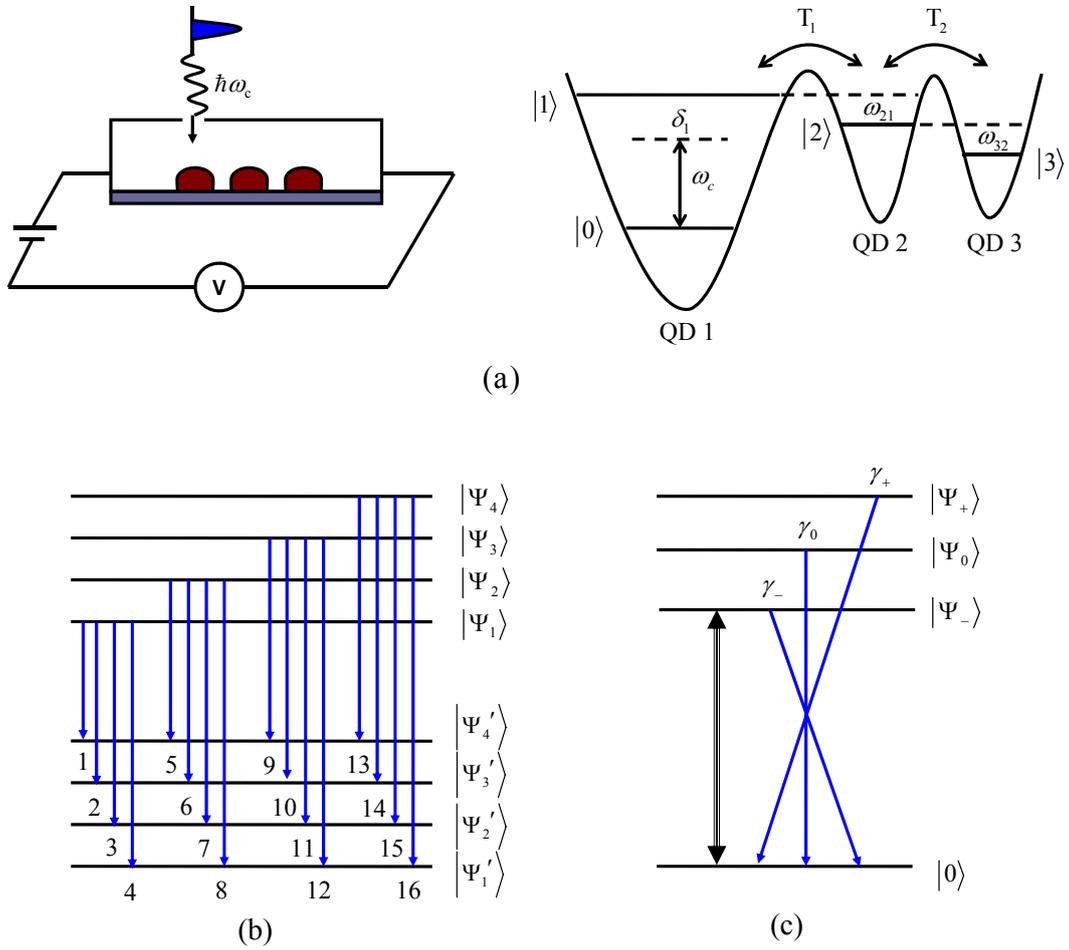

(a)

(b)

(c)

Fig. 1

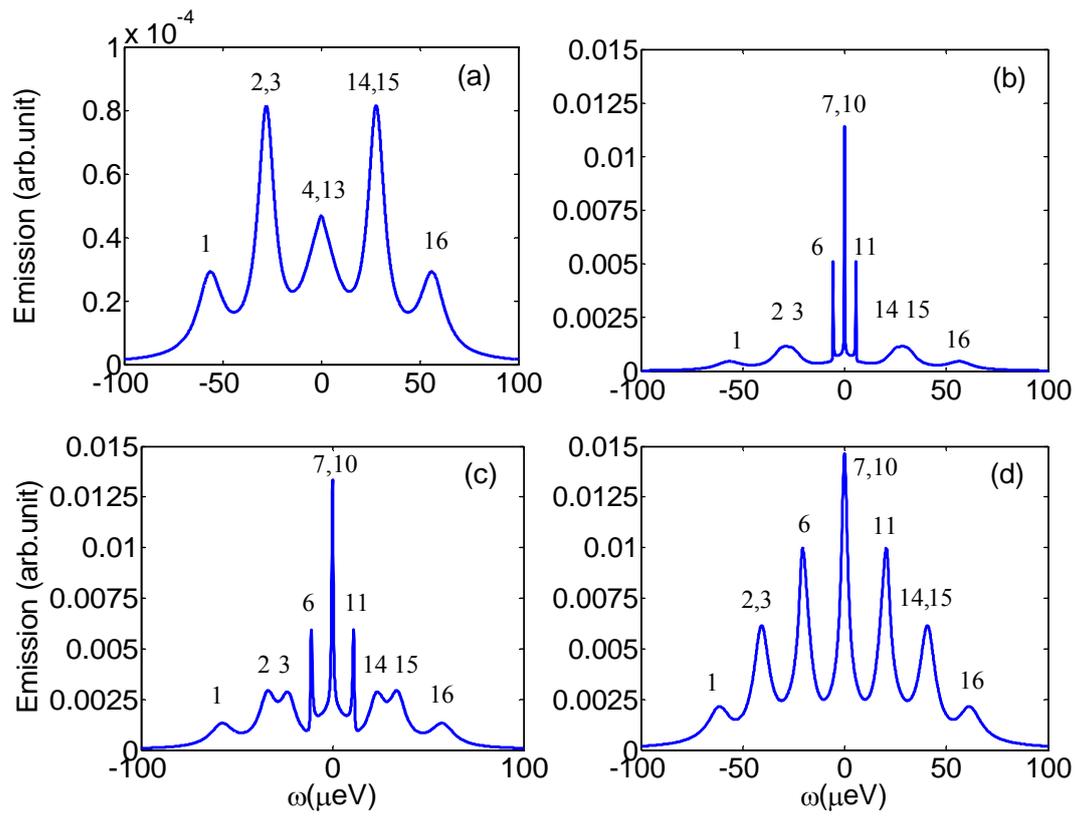

Fig. 2

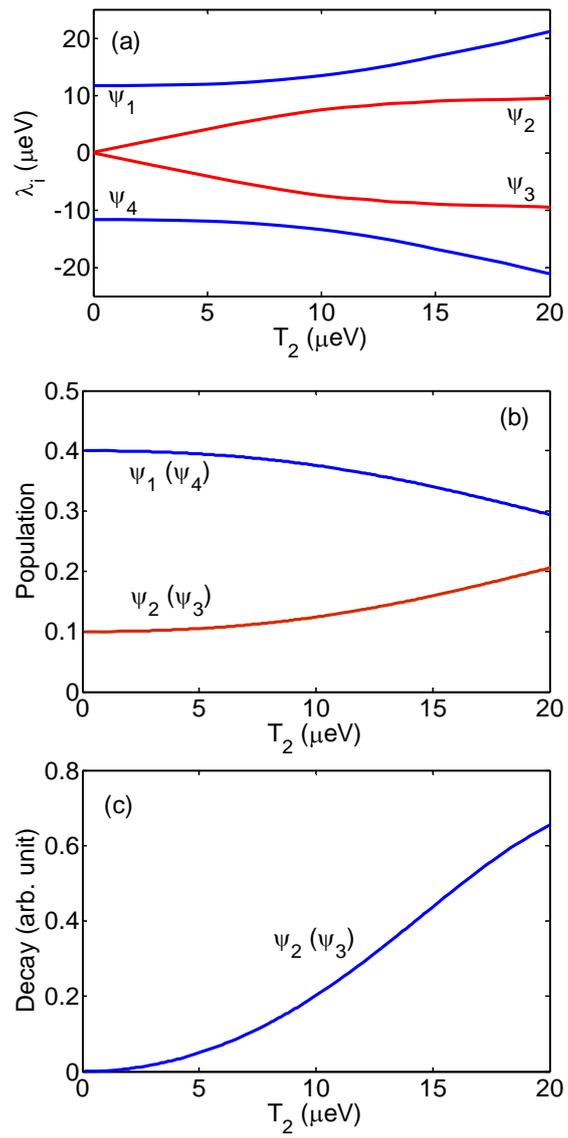

Fig. 3

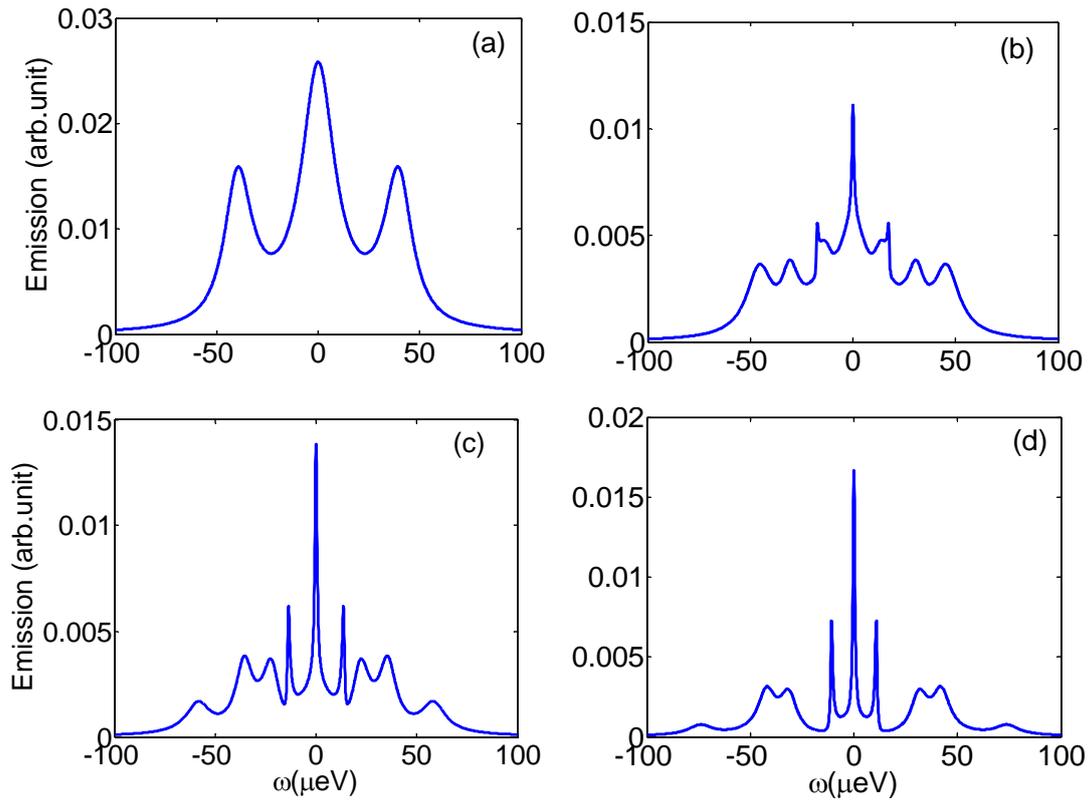

Fig. 4

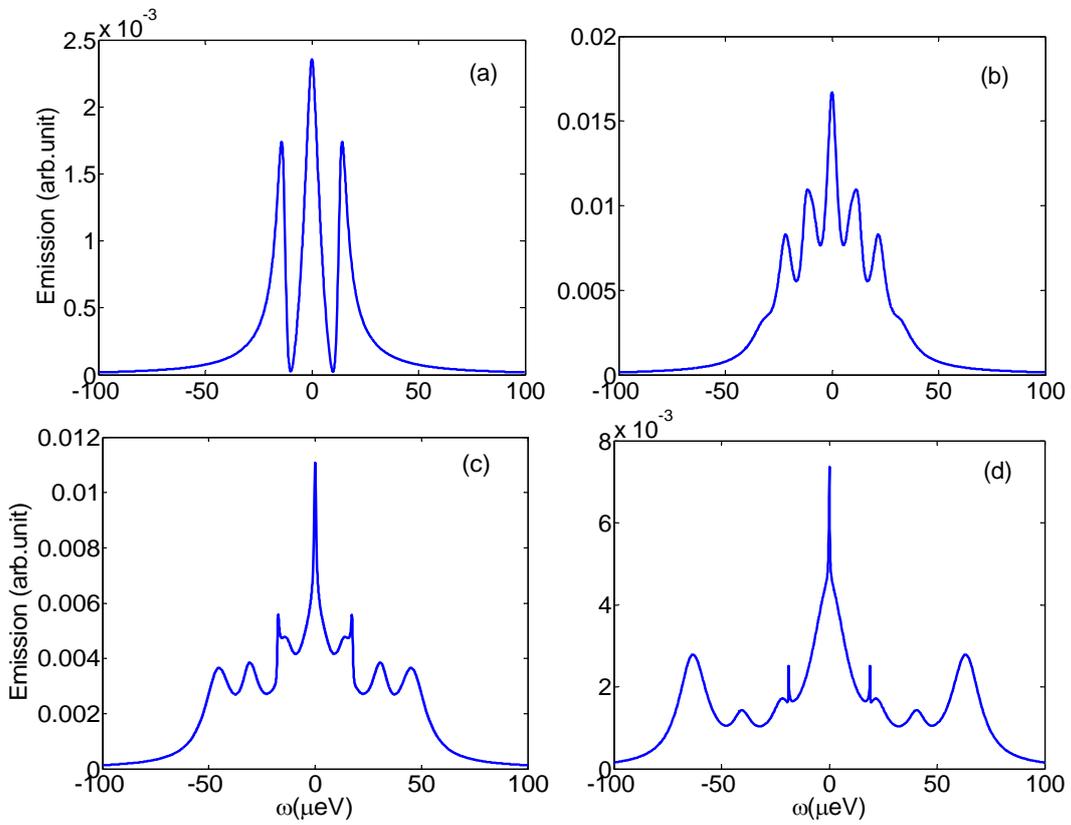

Fig.5

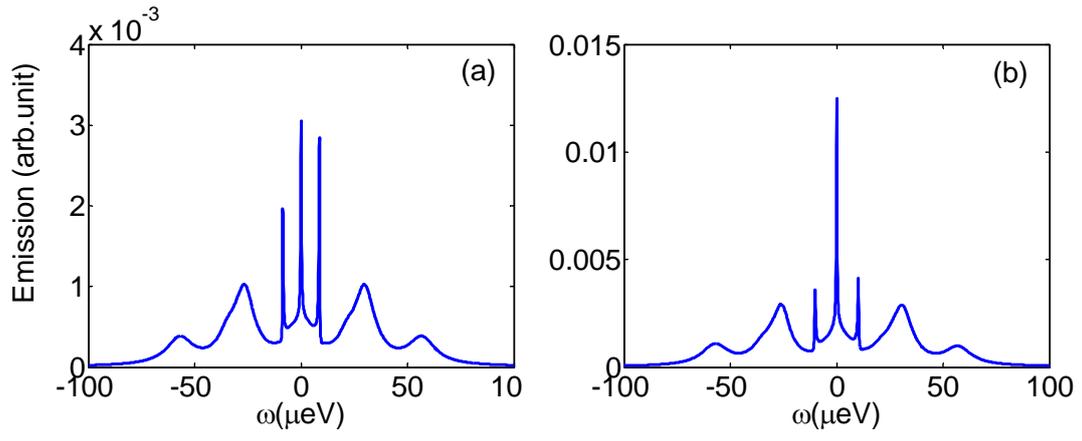

Fig. 6